\documentclass[%
reprint,twocolumn,jcp,superscriptaddress,showkeys,floatfix
]{revtex4-1}

\usepackage{graphicx}% Include figure files
\usepackage{dcolumn}% Align table columns on decimal point
\usepackage{bm}% bold math
\usepackage{amssymb}
\usepackage{amsmath}
\usepackage{color}
\usepackage[resetlabels]{multibib}
\usepackage{ifthen}

\newcommand{\dir}{.}

\newcommand{\etal}{{\em et al.\ }}
\newcommand{\via}{{\em via }}
\newcommand{\ie}{{\em i.e.\ }}
\newcommand{\eg}{{\em e.g.\ }}

\newcommand{\ud}{{\rm d}}

\newcommand{\DP}{{\rm DP}}
\newcommand{\DB}{{\rm DB}}
\newcommand{\ANB}{{\rm ANB}}
 
\newcommand{\new}[1]{\textcolor{black}{#1}}

\begin{document}

\title{Statistical properties of linear-hyperbranched graft copolymers 
prepared via "hypergrafting" of AB$_m$ monomers from linear 
B-functional core chains: A Molecular Dynamics simulation}
%\thanks{A footnote to the article title}%

\author{Hauke Rabbel}
\affiliation{Leibniz-Institut f\"ur Polymerforschung Dresden
    e.V., Hohe Stra\ss e 6, 01069 Dresden}

\author{Holger Frey}
\affiliation{Institut f\"ur Organische Chemie, 
Johannes Gutenberg-Universit\"at Mainz, 
D-55099 Mainz, Germany}

\author{Friederike Schmid} \email{friederike.schmid@uni-mainz.de}
\affiliation{Institut f\"ur Physik, Johannes Gutenberg-Universit\"at Mainz, 
D-55099 Mainz, Germany}

%\collaboration{MUSO Collaboration}%\noaffiliation

%\author{Charlie Author}
% \homepage{http://www.Second.institution.edu/~Charlie.Author}
%\affiliation{
% Second institution and/or address\\
% This line break forced% with \\
%}%
%\affiliation{
% Third institution, the second for Charlie Author
%}%
%\author{Delta Author}
%\affiliation{%
% Authors' institution and/or address\\
% This line break forced with \textbackslash\textbackslash
%}%

%\collaboration{CLEO Collaboration}%\noaffiliation

\date{\today}% It is always \today, today,
             %  but any date may be explicitly specified

\begin{abstract}

The reaction of AB${}_m$ monomers ($m=2,3$) with a multifunctional B${}_f$-type
polymer chain (''hypergrafting'') is studied by coarse-grained molecular
dynamics simulations. The AB${}_m$ monomers are hypergrafted using the slow
monomer addition strategy.  Fully dendronized, i.e., perfectly branched
polymers are also simulated for comparison. The degree of branching DB of the
molecules obtained with the ''hypergrafting' process critically depends on the
rate with which monomers attach to inner monomers compared to terminal
monomers. This ratio is more favorable if the AB${}_m$ monomers have lower
reactivity, since the free monomers then have time to diffuse inside the chain.
Configurational chain properties are also determined, showing that the
stretching of the polymer backbone as a consequence of the ''hypergrafting''
procedure is much less pronounced than for perfectly dendronized chains.
Furthermore, we analyze the scaling of various quantities with molecular weight
$M$ for large $M$ ($M>100$).  The Wiener index scales as $M^{2.3}$, which is
intermediate between linear chains ($M^3$) and perfectly branched polymers
($M^2 \ln(M)$). The polymer size, characterized by the radius of gyration $R_g$
or the hydrodynamic radius $R_h$, is found to scale as $R_{g,h} \propto
M^{\nu}$ with $\nu \approx 0.38$, which lies between the exponent of diffusion
limited aggregation ($\nu = 0.4$) and the mean-field exponent predicted by
Konkolewicz and coworkers ($\nu = 0.33$).  

\end{abstract}

\maketitle

\section{Introduction} 

Hyperbranched polymers are macromolecules with an irregular tree-like structure
\cite{frey_book}. 
Their dense structure provides them with unique
  mechanical and rheological properties, and the large number of
  functional end groups makes them interesting for applications in
  nanomedicine and material science \cite{gao04,voit09,irfan10,schuell13rev}.
Compared to dendrimers with a regular dendritic structure
  \cite{bosman99, grayson01, gillies05, klos09}, their synthesis is much 
  simpler and can often be done in just one reaction step \cite{schuell13rev}.
One particularly promising strategy for synthesizing hyperbranched
polymers is "hypergrafting", a "grafting-from" approach where AB${}_m$
monomers attach to a multifunctional macroinitiator core by slow monomer 
addition (SMA) 
\cite{schuell13rev,radke98,hanselmann98,bharathi00,moeck01,schuell12,schuell13}.
 The SMA technique allows to control the molecular weight to
some extent, polydispersities are comparatively low, and it minimizes side
reactions such as the formation of oligomers and cyclic side products.

Theoretical interest in hyperbranched polymers goes back to the early
Fifties of the last century \cite{zimm49,flory52}. The polycondensation of
AB${}_m$ monomers was studied by rate equations
\cite{radke98,hanselmann98,schuell13}, mean-field methods \cite{lubensky79,
isaacson80,stauffer82,cates85,vilgis88, vilgis92,buzza04,
konkolewicz07,konkolewicz08,konkolewicz11} and by computer simulations 
of lattice \cite{konkolewicz08, richards07, wang10, wang11,
juriju14} \new{and off-lattice models \cite{lescanec90}}.
Simple rate equation models predict that the polydispersity index
of the resulting polymers can be reduced by increasing the core functionality
\cite{radke98,hanselmann98,schuell13}.  Mean-field approaches have indicated
that the polymers have a self-similar architecture \cite{zimm49, lubensky79,
isaacson80, stauffer82, cates85, vilgis88, vilgis92, buzza04}. The radius of
gyration was predicted to scale with the chain length according to a power law
$R_g \sim N^{\nu}$ with Flory exponent varying between $\nu = 1/4$
\cite{zimm49} and $\nu = 1/2$ \cite{isaacson80}, depending on the
dimensionality \cite{lubensky79} and the theoretical approach
\cite{stauffer82}. In a relatively recent series of studies, Konkolewicz \etal
calculated the scaling behavior of molecules created by slow monomer addition
within a mean field theory that takes into account the evolution of the monomer
density profile of the chains during the growth process \cite{konkolewicz07,
konkolewicz10, konkolewicz08,konkolewicz11}. They predicted an initial
logarithmic scaling for small molecules, followed by a power law scaling with
exponent $\nu=1/3$ for larger molecules. They also tested their theory against
experiments and simulations and found good agreement
\cite{konkolewicz08,konkolewicz11,konkolewicz07b,konkolewicz10b}. 

Other recent simulation studies on the generation of hyperbranched polymers 
were presented by Wang \etal \cite{wang10,wang11} and by Juriju \etal
\cite{juriju14}.  When comparing simulations with each other, it is important
to specify the way how the hyperbranched chains were generated. In
the simulations of Konkolewicz \etal \cite{konkolewicz08}, random 
self-avoiding walks with lengths chosen from a prescribed distribution were 
grown one after the other, each starting from a randomly selected branching 
point.  In contrast, Wang \etal \cite{wang11} simulated a process of slow 
monomer addition under conditions of diffusion-controlled polymerization 
(\eg, a polymerization via radical reactions), where the monomers diffuse 
into the chain from outside and attach to the first reaction site they 
encounter. They report Flory exponents between 1/2 and 1/3. For large 
molecular weights, $\nu$ is found to approach the characteristic  exponent 
of diffusion limited aggregation (DLA), $\nu = 2/5$. Juriju \etal 
\cite{juriju14} examined two types of growth processes and identified two 
distinct universality classes associated with each of them. The first 
is a "quick growth" process which allows for cluster-cluster aggregation. 
This generates chains with a fractal structure and Flory exponent 
$\nu \approx 1/4$. The second is a "slow growth" process where monomers 
are attached sequentially to random sites of the chain, however, without 
having to diffuse there first. This was found to result in dense chains 
where $R_g$ scales logarithmically with the chain length, in accordance 
with the theoretical prediction of Konkolewicz for short chains 
\cite{konkolewicz07}. Juriju \etal \cite{juriju14} also investigated the 
effect of excluded volume interactions and found that they have almost 
no influence on the results.

\begin{figure}
\includegraphics[width=0.45\textwidth]
  {\dir/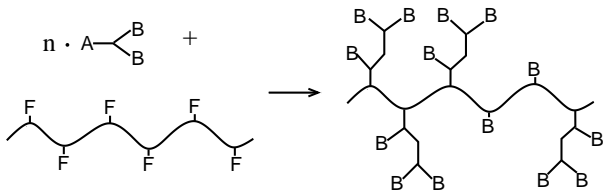}
\caption{\new{
Schematic example of hypergrafting of AB${}_2$ monomers on a 
linear ($F_f$) backbone chain. }}
\label{fig:cartoon}
\end{figure}

In the present paper, we focus on linear-hyperbranched graft copolymers
(LHGCs), which have recently been proposed as an interesting alternative to
standard hyperbranched polymers and dendronized polymers. Dendronized polymers
have a linear backbone decorated with dendritic side chains \cite{schlueter00,
zhang03, frauenrath05}, which gives them a cylindrical shape. While they are
interesting macromolecular objects, their production is as cumbersome as that
of dendrimers. LHGCs on the other hand also have a linear backbone, but use
hyperbranched side chains instead
\cite{schuell13rev,schuell12,lach98,sun12,schuell12b}. They can be
synthesized by hypergrafting from the backbone (see Fig.\ \ref{fig:cartoon}).
In a recent paper \cite{schuell13}, we have calculated the expected
polydispersity index of such chains, taking into account the core dispersity,
within a rate equation theory.  Here we will present off-lattice Monte Carlos
simulation of the growth of such chains under slow monomer addition, assuming
that the polymerization reaction is diffusion controlled, and analyze the
resulting topological and conformational properties.

The paper is organized as follows. In the next section, we introduce the
simulation model and method. Then we will discuss the properties of the
resulting chains, focussing on statistical and topological properties in
section \ref{sec:statistical} and on conformational properties in section
\ref{sec:configurational}. We summarize and conclude in section
\ref{sec:conclusions}.

\section{Simulation Model}
\label{sec:model}

We use a simple bead-spring model for the chains. Beads repel each other \via a
repulsive WCA potential \cite{wca} with diameter $\sigma$ and energy prefactor
$\epsilon$, and are connected by harmonic bonds with spring constant $k=10
\epsilon/\sigma^2$, resulting in the equilibrium distance $r_0 \approx 1.1
\sigma$. In addition, a bending potential $V_b = K(1-\cos(\Phi))$ is applied,
where $\Phi$ is the angle between subsequent bonds and the bending constant $K$
is chosen $K=10 \epsilon$. Here and throughout, the simulation units of length
and energy are the interaction parameter $\sigma$ and $\epsilon$ of the WCA
potential, and the time unit is $\tau = \sqrt{m \sigma/\epsilon}$, where $m$ is
the mass of one bead. AB${}_m$ monomers are modelled as dimers of two beads,
one bead representing the A reactive group, and the other one all the B groups.
We do not include explicit solvent. Since the beads repel each other,
our system corresponds to a polymer in solvent under good solvent conditions. 
We carry out Molecular Dynamics simulations using the Velocity-Verlet algorithm
and a Langevin thermostat at temperature $k_B T = 0.5 \epsilon$ with the time
step $0.005 \tau$. The simulations were carried out using the open source
program package ESPResSo \cite{espresso}, which we extended to allow for 
chemical reactions between monomers and the main chain. 

The monomers are created and equilibrated in a separate reservoir and then
added one by one in intervals of $1000 \tau$ to the main system, which
initially contains  one equilibrated backbone chain made of $f$ beads
\new{F}.  This is done  by placing the monomers randomly on the surface of
a sphere centered at the center of mass of the polymer, whose radius is $5
\sigma$ larger than the largest distance between the center and a chain bead
(A, B, or \new{F}). Fully periodic boundary conditions are applied such
that monomers cannot leave the simulation box. Once they have diffused close to
a chain bead, they may react with the chain, thus making the polymer grow. In
that case, a new harmonic bond is established between an A bead and a B or
\new{F} bead. 

Reactions may take place under the following conditions
\begin{itemize}
\item The distance of the potential reaction partners ((A,B) or 
(A,\new{F})) is less than a critical distance $r_c=1.1 \sigma$.
\item None of the beads involved in a reaction have already reached
the maximum number of allowed reactions. A and \new{F} beads can react 
only once, B beads may react $m$ times.
\item One of the partners belongs to the central chain. Reactions
between free AB${}_m$ monomers are not allowed. 
\end{itemize} 

If these conditions are fulfilled in a given time step, the monomers react with
each other with a probability $p$. We chose $p=1$ for reactions with the
backbone chain \new{((A,F) reactions)}, and $p=1$ ("high reactivity") or
$p=0.01$ ("low reactivity") for (A,B) reactions.  Since monomer pairs that meet
the above conditions in a given time step may still have a distance less than
$r_c$ in the subsequent time steps, the net reactivity (\ie, the fraction of
collisions that lead to bond formation) is not equal  to $p$ for $p<1$. In our
simulations, we found that $p=0.01$ results in a net reactivity of roughly $r
\approx 1/3$.  (For $p=1$, the reactivity is obviously $r=1$.) \new{In the
present work, we take $p$ to be the same for AB${}2$ and AB${}_3$ monomers.
However, since the monomers also have to move close to each other before they
can react, the resulting reaction rates may be different for AB${}_2$ and
AB${}_3$ monomers as described below.} 

\section{Results}
\label{sec:results}

We will now present the simulation results, focussing first on the topological
properties of the chains (\ie, their architecture), and then on the resulting
configurational properties (radius of gyration etc.). 

\begin{figure}
\includegraphics[width=0.45\textwidth]{\dir/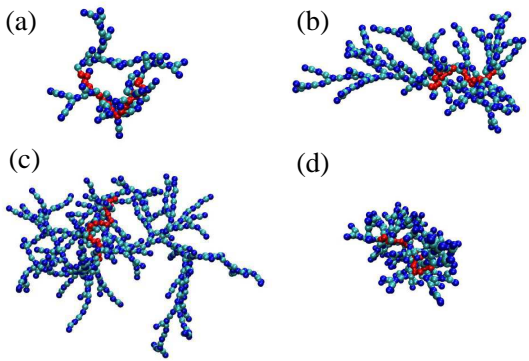}
\caption{
Simulation snapshot of hyperbranched polymers with backbone length $f=20$ and
monomer functionality $m=2$. The degrees of polymerization (i.e.\ the number of
bound monomers) are (a) $\DP$=60, (b)  $\DP$=140, (c) $\DP$=290. (d) shows a
perfectly dendronized polymer of generation 3 on the same backbone 
for comparison with the hyperbranched structures ($\DP$=140).}
\label{fig:snapshots}
\end{figure}

For every choice of backbone functionality $f$ ($f=20,40,100$) and monomer
functionality $m$ ($m=2,3$), 20-25 independent simulation runs were carried
out. For comparison, we have also simulated fully dendronized polymers 
with backbone functionalities $f=20,40,100$ and monomer functionalities $m=2,3$.
Typical snapshots of hyperbranched polymers and one dendronized polymer are
shown in Fig.\ \ref{fig:snapshots}. 
  
\subsection{Statistical and topological properties of chains}
\label{sec:statistical}

We first address the important question how efficiently the slow monomer
addition process produces highly branched molecules, \ie, how close
the resulting molecular structures are to perfectly branched, dendronized
polymers.  To this end, we consider
\begin{itemize}
\item The backbone conversion, \ie, the average number of backbone groups
that have reacted with a monomer
\item the ''degree of branching'' \cite{hoelter97,hoelter97b}, which quantifies 
the efficiency with which the full branching capabilities of monomers 
are exploited in the process.
\end{itemize}

\begin{figure}
\includegraphics[angle=0, width=0.45\textwidth]
  {\dir/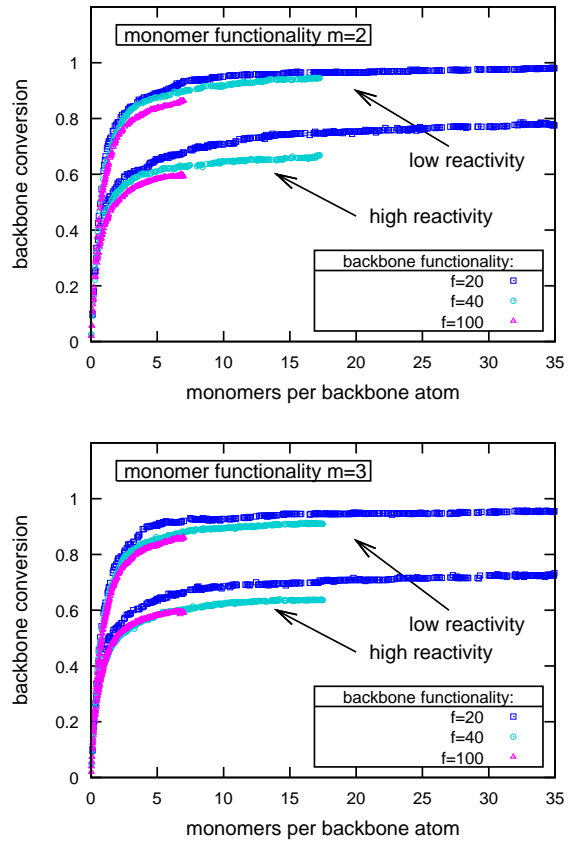}
\caption{
Mean backbone conversion as a function of the monomers per backbone unit
for monomer functionality $m=2$ (top) and $m=3$ (bottom), different
reactivities $r=1$ (''high'') and $r \approx 1/3$ (''low''), and different
backbone functionalities $f$ as indicated. The error bars (not shown) are
about twice the symbol size.}
\label{fig:backbone_conversion}
\end{figure}

The backbone conversion is shown in Fig.\  \ref{fig:backbone_conversion} as a
function of the grafted monomers per backbone unit in the chain - which is in
some sense a time axis in the polymerization simulation.  The curves for
AB${}_2$ and AB${}_3$ monomers are very similar. In the initial regime, where
monomers mainly react with backbone units, the backbone conversion increases
linearly. Then, as monomers start interacting with already formed
branching points, the curves deviate from linear growth and level off.  For
monomers with low reactivity, almost full conversion can be reached in the
course of the simulation. For highly reactive monomers, a substantial fraction
($\sim 20 \%$) of backbone reactive sites remain empty.

\begin{figure}
\includegraphics[angle=0, width=0.45\textwidth]
  {\dir/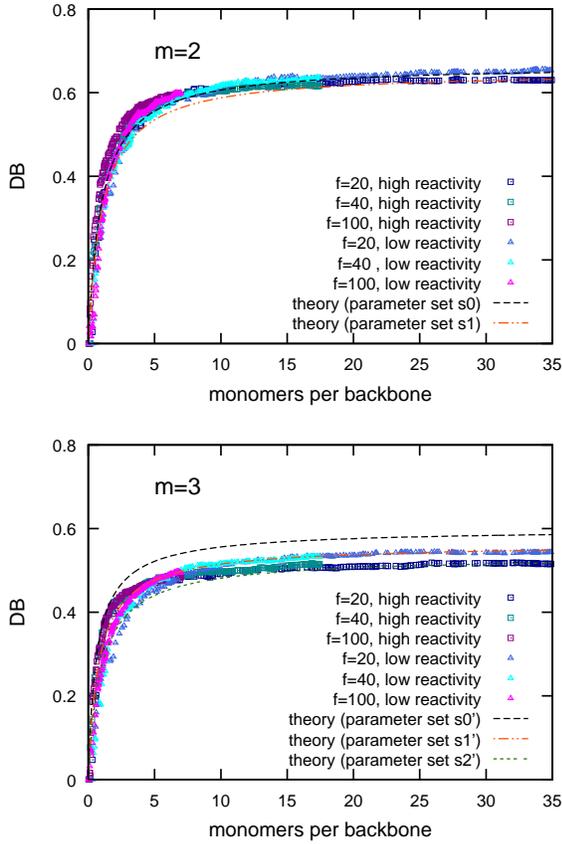}
\caption{
Mean $\DB$ as a function of monomers per backbone unit for monomer
functionalities $m=2$ (top) and $m=3$ (bottom), different backbone
functionalities $f$ as indicated, and different reactivities. 
Lines indicate theoretical predictions for different sets of
rate parameters
(cf.\ Eqs.\ (\protect\ref{eq:rate_1}-\ref{eq:rate_3})):
$k_r \propto (m-r)$ (s0 and s0');
$k_1/k_0 = 0.45$ (s1); $k_1/k_0 = 0.5, k_2/k_0 = 0.3$ (s1');
$k_1/k_0 = 0.45, k_2/k_0 = 0.27$ (s2'), and $k_b=k_0/m$ (all sets).
Sets s0,s1,s1', and s2' are taken from Fig.\ \protect\ref{fig:rates}.
}
\label{fig:DB}
\end{figure}

\begin{figure}
\includegraphics[angle=0, width=0.45\textwidth]
  {\dir/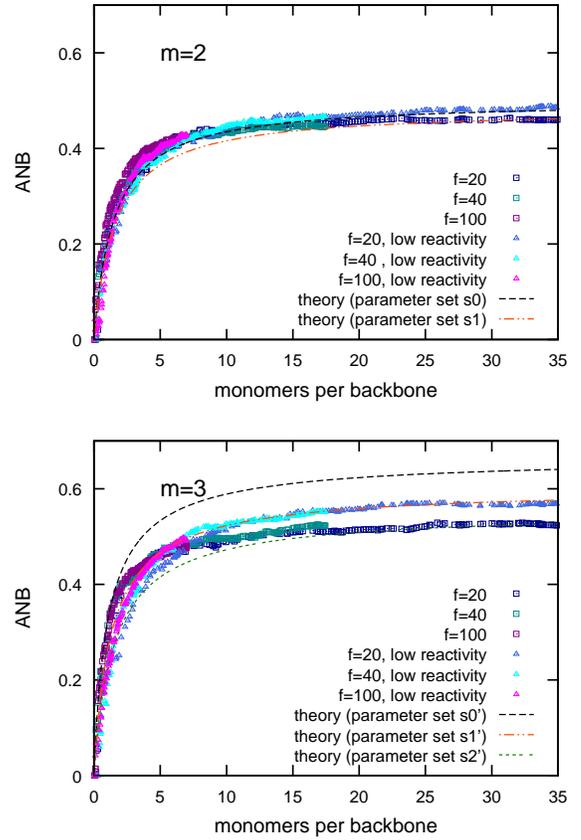}
\caption{
Mean average number of side branches $\ANB$ as a function of
monomers per backbone for monomer
functionalities $m=2$ (top) and $m=3$ (bottom), different backbone
functionalities $f$ as indicated, and different reactivities. 
Lines indicate theoretical predictions for the same parameter
sets as in Fig.\ \protect\ref{fig:DB}. 
}
\label{fig:ANB}
\end{figure}

The degree of branching can be quantified by a measure proposed by
H\"olter and Frey \cite{hoelter97}
\begin{equation}
\DB = \frac{m}{m-1} \: \frac{\sum_{r=1}^m (r-1) \: N_r}{\sum_{r=1}^m r \: N_r},
\end{equation}
where $N_r$ is the number of monomers in a chain that serve as starting point 
for $r$ branches. (Hence $N_0$ is the number of terminal units, $N_1$ the 
number of linear units etc.). $\DB$ quantifies the number of side branches
in the molecule relative to the total number of branches and is normalized
such that it varies between 0 (linear chains) and 1 (perfectly branched
dendronized chains). It is shown as a function of monomer units per backbone
unit in Fig.\  \ref{fig:DB}. The curves look similar to those for the
backbone conversion (Fig.\ \ref{fig:backbone_conversion}): They first
increase and then level off. They are higher for AB${}_2$ monomers than
for AB${}_3$ monomers, implying that it is less likely for AB${}_3$
than for AB${}_2$ monomers to fill all reaction sites. Nevertheless,
the total number of side branches per non-terminal monomer, quantified as
\begin{equation}
\ANB = \frac{\sum_{r=1}^m (r-1) N_r}{\sum_{r=1}^m N_r}
\end{equation}
is higher for AB${}_3$ monomers than for AB${}_2$ monomers as one
would expect. This is shown in Fig.\ \ref{fig:ANB}. Both DB and ANB 
increase slightly for less reactive monomers, but the influence of
reactivity on DB and ANB is much smaller than on the backbone conversion.

The behavior of the degree of branching $\DB$ and the average
number of side branches $\ANB$ can be rationalized within a simple 
rate equation approach proposed in Ref.\ \cite{hoelter97b}, which we slightly
modify to account for the effect of backbone conversion. The resulting
equations read

\begin{eqnarray}
\label{eq:rate_1}
\dot{\new{N_F}} &=& - c k_b \new{N_F} \\
\label{eq:rate_2}
\dot{N}_0 &=& c \big[ 
  k_b \new{N_F} + \sum_{r=1}^m \: k_r \: N_r  \big] \\
\label{eq:rate_3}
\dot{N}_r &=& c \big[
  k_{r-1} \: N_{r-1} - \: k_r \: N_r \big]: \;
  1 \le r \le m
\end{eqnarray}

Here $c$ is the concentration of unreacted monomers in solution, $N_B$ 
the number of unreacted backbone sites, 
and $k_r$ gives the rate at which a free monomer reacts with a monomer
that has already $r$ branches. Using these equations, one can 
calculate $\DB$ and $\ANB$ numerically as a function of time and 
derive the limiting behavior for large molecules. 

\begin{figure}
\includegraphics[angle=0, width=0.45\textwidth]
  {\dir/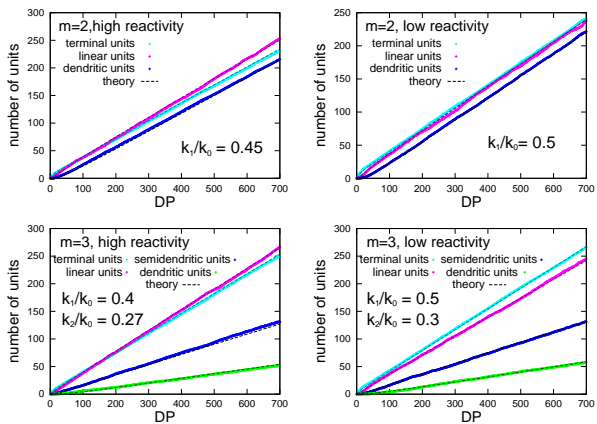}
\caption{
Average number of terminal ($N_0$), linear ($N_1$), 
dendritic ($N_m$) and semidendritic (for AB${}_3$ monomers: $N_2$) 
units in the chains as a function of the degree
of polymerization $\DP$; 
(a) $m=2$, high reactivity, 
(b) $m=2$, low reactivity, 
(c) $m=3$, high reactivity, 
(d) $m=3$, low reactivity. 
The backbone functionality in this example is $f=20$.
The evolution of the system is compatible with the effective
rate theory, Eqs.\ (\protect\ref{eq:rate_1})--(\protect\ref{eq:rate_3})
(black dashed lines) with rate parameters as given in the figures
and $k_b = k_0/m$ (the results are not sensitive to the choice of $k_b$).
}
\label{fig:rates}
\end{figure}

In the late stages in the polymerization process, the contribution of $B$ is
negligible and the relative fraction of all other units reaches a steady state,
\ie, $\dot{N}_r/\sum_r \dot{N_r} \propto N_r/\sum_r N_r$, where 
$\sum_{r=0}^m N_r = \DP$ is the degree of polymerization.
The solution becomes particularly simple if the reaction rates take the
simple intuitive form $k_r \propto (m-r) \hat{k}$, where the
combinatorial factor ($m-r$) accounts for the number of remaining available
reaction sites in an AB${}_m$ monomer that has already reacted with $r$
monomers. In that case, one obtains the limiting behavior $\DB \to
\frac{m}{2m-1}$ and $\ANB \to \frac{m-1}{m}$ \cite{hoelter97b}.  In our
simulation model, we cannot expect 
$k_r \propto (m-r) \hat{k}$ to hold, since the
''reaction sites'' are on the same bead. However, we can verify whether $N_r$ 
is a linear function of $\DP$ at late stages of the polymerization process
as predicted, and Fig.\ \ref{fig:rates} shows that this is indeed the case. 

Extracting the rate constants $k_r$ of the theory from these data, we can then
calculate $\DB$ and $\ANB$ as a function of the degree of polymerization per
backbone ($\DP/f$).  This gives the lines shown in Figs.\ \ref{fig:DB} and
\ref{fig:ANB}, which are in very good agreement with the simulation data except
in the initial polzmerization process where molecules are still very small.

We conclude that the simulation data can be described very well by the
simple rate theory.  One consequence is that the degree of branching of 
the molecules obtained with this polymerization process mostly depends on 
the ratio $k_r/k_0$ of rates with which monomers attach to inner monomers 
compared to terminal monomers. This ratio is more favorable if monomers have 
lower reactivity, since the free monomers then have time to diffuse inside 
the chain and are not captured by the first (usually terminal) chain segments 
they encounter. For AB${}_3$ monomers, the rate $k_2/k_0$ is less 
favorable than the rate $k_1/k_0$. This is a geometrical effect: Once a 
monomer has grown one side branch, it is less accessible for free monomers 
and the probability to grow a second side branch drops.

\begin{figure}
\includegraphics[angle=0, width=0.45\textwidth]
  {\dir/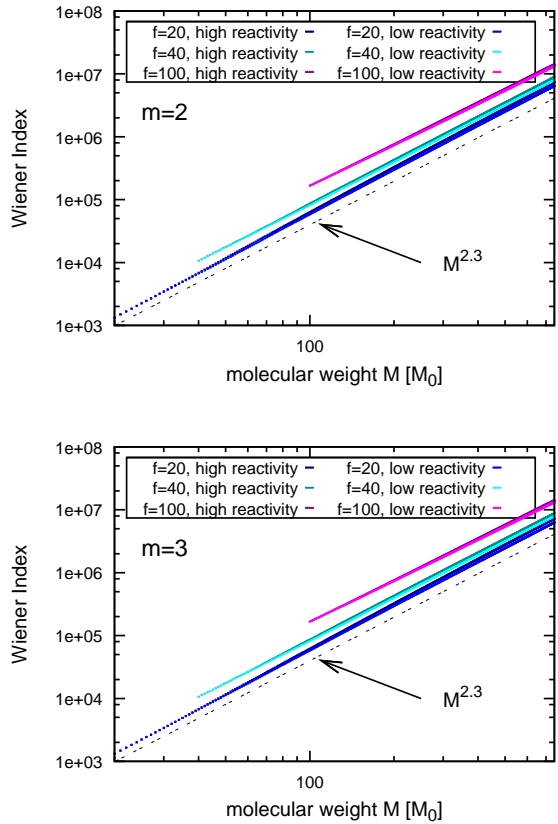}
\caption{
Wiener index of the hyperbranched polymers as a function of the molecular
weight $M$ in units of monomer mass $M_0$ for monomer functionality $m=2$ (top)
and $m=3$ (bottom) and different core functionalities $f$ as indicated. At 
high molecular weight, the Wiener index displays a power law behavior.
}
\label{fig:wiener}
\end{figure}

\begin{figure}
\includegraphics[angle=0, width=0.45\textwidth]
  {\dir/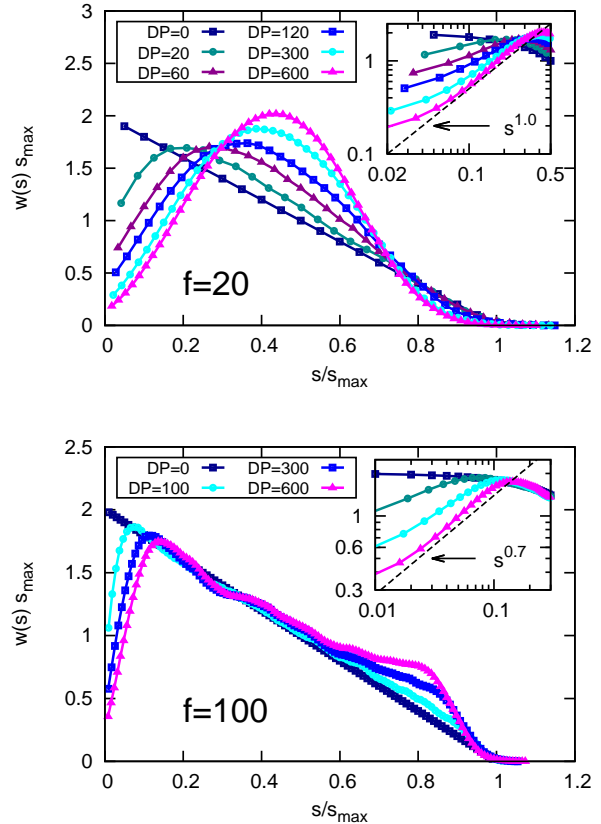}
\caption{
\new{
Averaged distribution of strand lengths of the hyperbranched polymers for
backbone length $f=20$ (top) and $f=100$ (bottom) at different degrees of
polymerization $DP$ as indicated. $DP=0$ corresponds to the bare backbone, with
the distribution $w(s)$ described by Eqn.\ \ref{eq:strandlength_dist_linear}.
The plots show the data from simulations with high reactivity. Insets show the
low $s/s_{max}$ regime of the same data on a double logarithmic scale, along
with a power law function (black line) to give an idea of the behavior. Here,
$s_{max}$ is the average length of the longest strand. 
}
}
\label{fig:strandlength_dist}
\end{figure}

Apart from the degree of branching, the Wiener index $W(T)$ is another quantity
that can be used to describe the topology of a branched structure with the
topology of a tree $T$. It was first introduced and applied in a chemical 
context by Harry Wiener \cite{wiener47}, but has numerous applications in 
mathematics and other fields, too \cite{dobrynin01}.  For a tree consisting 
of $N+1$ vertices (in our case monomers) connected by $N$ edges (bonds created
by reactions), the Wiener index is defined as the sum over all ''distances'' 
$s(v_i,v_j)$ between any two vertices $v_i$ and $v_j$,
\begin{equation}
  \label{eq:def_wienerindex}
  W=\frac{1}{2}\sum_{ij}s(v_i,v_j).
\end{equation}
Here the ''distances'' $s(v_i,v_j)$ are taken along the edges of the tree, and
the length of one edge equals 1.  
%The Wiener index normalized to the number of
%pairs of vertices in the tree $N(N-1)/2$ is also known as the average (total)
%path length $P(T)$ (equation \ref{eq:average_path_length}).
%\begin{equation}
%  \label{eq:average_path_length}
%  p(T)=\frac{2 W}{N(N-1)}=\frac{1}{N(N-1)}\sum_{ij}d(v_i,v_j)
%\end{equation}
The Wiener index and the path lengths can be used to compare densities of
trees. Even though it can not be measured experimentally, it has been shown to
correlate reasonably with properties such as density, viscosity and melting
point \cite{bonchev92,widmann98,sheridan02}. For a linear chain
consisting of $N+1$ vertices, the Wiener index is given by
$W=\frac{1}{6} N (N+1) (N+2)$ and hence scales as $W \sim N^3$ 
\cite{canfield85}. It becomes smaller for branched molecules. 
For \new{fully} dendronized molecules, it scales as $W \sim N^2 \ln(N)$ 
\cite{note,gutman93}. 
\new{$W$ can also be related to the first moment of the
  distribution $w(s)$ of distances or strand lengths $s$ \cite{polinska2014}. 
  With $w(s)$ normalized as
\begin{equation}
  \label{eq:strandlength_dist}
  w(s)=\frac{1}{N^2}\sum_{i,j}\delta(s-s_{ij}),
\end{equation} 
the Wiener index can be calculated according to 
\mbox{$W=\frac{1}{2}N^2\sum s\,w(s)$}. For a linear chain
$w(s)$ is given by
\begin{equation}
  \label{eq:strandlength_dist_linear}
  w(s)=\frac{2}{s_{max}}\left(1-\frac{s}{s_{max}}\right)
\end{equation} 
where $s_{max}=N-1\approx N$ is the longest strand.
}
\new{
The simulation results for these quantities are shown in 
Fig.\ \ref{fig:wiener} (Wiener index) and 
Fig.\ \ref{fig:strandlength_dist} (strand length distribution). 
}% closing new environment

We find that the Wiener index roughly scales as $W \sim M^{2.3}$ with the
molecular weight $M = f + \DP$, hence the scaling exponent is less than that of
linear chains, but higher than that expected for dendronized molecules.  The
monomer functionality $m$ does not have an influence on $W$, and the influence
of the monomer reactivity is also very small.
\new{
The distributions $w(s)$ are described by Eqn. 
(\ref{eq:strandlength_dist_linear}) for $DP=0$ (linear initiator chains). For
$DP>0$, $w(s)$ first displays an increase caused by the branched structures,
followed by a decrease due to the finite length of the strands. For self
similar structures, the initial increase should be described by a power law
\cite{polinska2014}. Such a behavior is not clearly identified here (insets
in Fig.\ \ref{fig:strandlength_dist}). This is not surprising, given the fact
that branched structures are grown on linear initiators, and that the
branched structures are not much larger than the backbone, especially for long
backbones $f=100$. Consistent with this, the decay of $w(s)$ after the initial 
increase is dominated by the linear chain behavior in this case. 
Interestingly, for $f=100$ at high strand lengths, the distributions 
increasingly deviate from the linear chain behavior
with growing degree of polymerization, and a small shoulder develops
at $s/s_{max} \approx 0.8$ (Fig.\ \ref{fig:strandlength_dist} (bottom)) . 
For comparison, we have also calculated the strand length distribution for
ideal hyperbranched chains, which were constructed based on the rate
Eqs.\ (\ref{eq:rate_1}) -- (\ref{eq:rate_3}) (data not shown). The
shoulder did not emerge there. We conclude that it is most likely 
caused by the inhomogeneous distribution of branched side chains 
along the backbone, which is analyzed below.
}%closing new environment

\begin{figure}
\includegraphics[angle=0, width=0.45\textwidth]
  {\dir/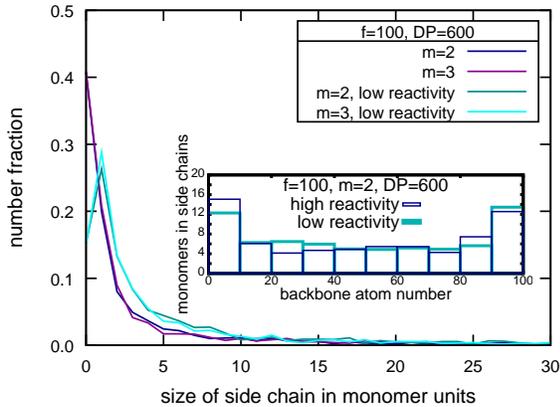}
\caption{
Normalized distribution of side chain masses for molecules with core 
functionality $f=100$, monomer functionality $m=2$, and total degree 
of polymerization $\DP=600$. One side chain hence contains 6 monomer
units on average.  Inset shows distribution of polymer mass along the backbone
of the chains. The units of the backbone chain are labelled 
from 0 (one end) to 100 (other end).
}
\label{fig:monomer_distribution}
\end{figure}

Last in this subsection, we study the distribution of monomers on the side
chains. The effect of monomer reactivity on the mass distribution is
illustrated in Fig.\ \ref{fig:monomer_distribution}, which shows the histogram
of masses of side chains made of AB${}_2$ monomers in molecule with backbone
functionality $f=100$ and total polymerization index $\DP=600$ .  For highly
reactive monomers, the curve decreases monotonically and rapidly to a value
close to zero. Since the average side chain length must be 6 monomers, this
implies that molecules must contain a few long side chains which compensate for
the many very short chains.  Decreasing the monomer reactivity reduces the
fluctuations and leads to some compactification.  The minimum of the
distribution is shifted away from zero and the number of side chains with 1-10
monomers increases.  The inset of Fig.\  \ref{fig:monomer_distribution} shows
the mass distribution (mass in monomer units) along the backbone in the same
chains.  One can clearly see that the mass accumulates at both ends of the
backbone. While the mass distribution is approximately uniform in the middle
region, the side chains that have grown from the ends of the backbone are about
twice as big on average.  This effect is observed both for highly reactive
monomers and for monomers with reduced reactivity. It can presumably
be explained by a reduction of screening close to the backbone ends.
Side branches grafted at the middle of the backbone chain are surrounded by
competing side branches that may also capture monomers, hence they are
less accessible for diffusing monomers than side branches grafted at the
ends of the backbone.

\subsection{Configurational properties}
\label{sec:configurational}

After having analyzed the topological properties of the linear-hyperbranched
chains generated by the hypergrafting process, we now study
the configurational properties. Specifically, we examine the 
the gyration radius and the hydrodynamic radius, and the amount of 
backbone stretching induced by hypergrafting side chains.

\new{For ideal Gaussian chains, the radius of gyration
\begin{equation}
R_g^2 = \frac{1}{2 N^2} \sum_{i,j=1}^N (\vec{r}_i - \vec{r}_j)^2
\end{equation}
(here the sum runs over all $N$ beads of the molecules) is directly
related to the Wiener index, Eq.\ (\ref{eq:def_wienerindex}),
{\em via} $\langle (\vec{r}_i - \vec{r}_j)^2 \rangle \propto s(v_i,v_j)$. 
Inserting our result $W \sim N^{2.3}$, this
would suggest $R_g \propto \sqrt{W}/N \sim N^{\nu}$ with $\nu = 0.15$.
However, as we shall see below, the actual exponent $\nu$ is much larger due 
to excluded volume effects.
}

\begin{figure}
\includegraphics[angle=0, width=0.45\textwidth] {\dir/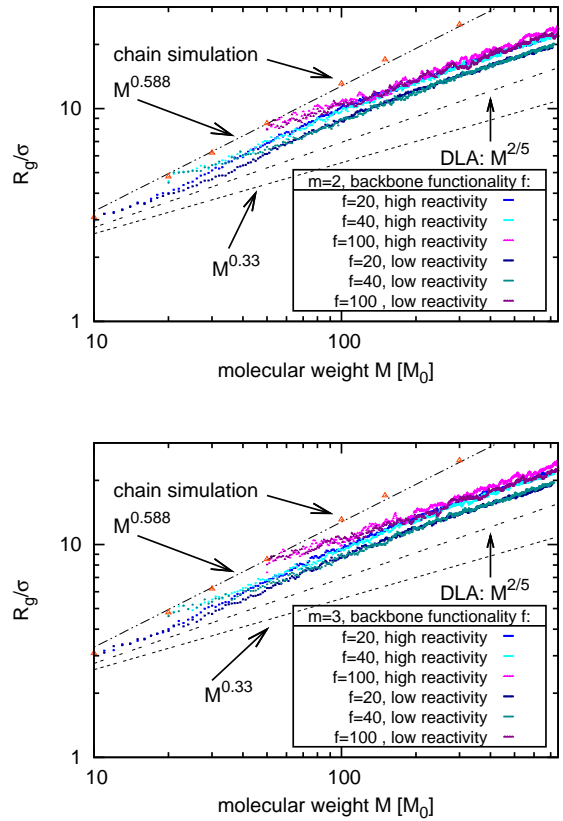}
\caption{
Radius of gyration versus molecular weight (in monomer units) for 
monomer functionality $m=2$ (top) and $m=3$ (bottom),
various backbone core functionalities $f$, and the two different choices of
reactivity. Also shown for comparison are the results from simulations
of linear chains. Dashed lines indicate slopes corresponding to 
different scaling behavior.
}
\label{fig:gyration_1}
\end{figure}

Fig.\ \ref{fig:gyration_1} summarizes our data for the radius of gyration $R_g$
as a function of molecular weight for all backbone functionalities $f$, monomer
functionalities $m$, and the two choices of monomer reactivity. 
%The radius of gyration was calculated using
%\begin{equation}
%R_g^2 = \frac{1}{N^2} \sum_{i,j=1}^N (\vec{r}_i - \vec{r}_j)^2,
%\end{equation}
%where the sum runs over all $N$ beads of the molecule.  
The data ware gathered ''on the fly'' during the simulation of the
polymerization reaction.  The results for $m=2$ and $m=3$ are essentially
identical. At late stages of the polymerization, all data collapse on two
curves, one for high reactivity and one for low reactivity.  The only
exceptions are the curves for $f=100$ and low monomer reactivity, which
apparently do not reach the asymptotic limit for the molecular weights under
consideration. In the limit of large molecular weight, the two curves have the
same slope in a double logarithmic plot.  The corresponding scaling exponent,
$\nu = 0.38(1)$, lies between the exponent of DLA scaling ($\nu = 0.4$) and the
exponent predicted by Konkolewicz \etal \cite{konkolewicz07} ($\nu = 0.33$),
and is clearly much smaller than the Flory exponent for linear chains ($\nu =
0.588$).

\begin{figure}
\includegraphics[angle=270, width=0.45\textwidth] 
  {\dir/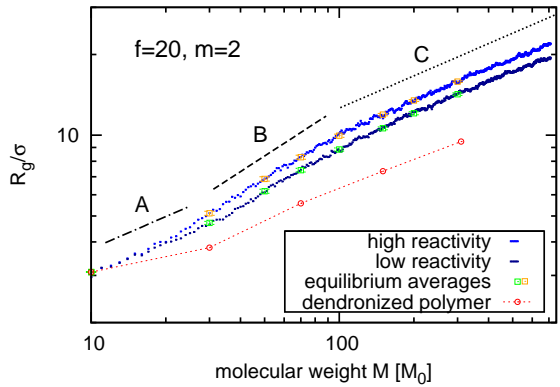}
\caption{
Radius of gyration versus molecular weight in monomer units for monomer
functionality $m=2$ and backbone functionality $f=20$ (top) and $f=40$
(bottom). Small symbols correspond to data that were collected during the
polymerization simulation, larger symbols to data that were obtained from
separate, longer simulations.  Also shown for comparison are data from
simulations of fully dendronized polymers.
}
\label{fig:gyration_2}
\end{figure}

Since the data shown in Fig.\ \ref{fig:gyration_1} were gathered in a
simulation of steadily growing chains, it is not clear whether they reflect the
properties of fully equilibrated chains. To clarify this point, we have stored
the architectures obtained for selected degrees of polymerization and studied
their configurational properties by separate equilibrium simulations without
monomer addition steps. The results for $R_g$ are shown in Fig.\
\ref{fig:gyration_2}. They do not differ noticeably from the data taken in the
polymerization simulation. Hence, we can conclude that the polymers in the
polymerization simulation are equilibrated despite the occasional addition of
one monomer. For comparison, we also show data for dendronized polymers. We
find that the latter are much more compact, and in addition, the scaling of
$R_g$ with the molecular weight seems different. Our data for the
perfectly branched dendrimers are compatible with previous dendrimer
simulations \cite{goetze03,giupponi04} that have suggested a scaling between
$R_g \propto M^{1/5}$ (the Flory prediction) and $R_g \propto M^{1/3}$. The
data for the hyperbranched chains suggest three different scaling regimes: An
initial regime (A) with a scaling exponent of roughly $\nu \approx 0.4$, a
second regime (B) where $R_g$ grows even stronger as a function of $M$ with an
exponent $\nu \approx 0.5$, and a third regime (C) where the exponent $\nu
\approx 0.4$ is recovered. 

\begin{figure}
\includegraphics[angle=0, width=0.45\textwidth] 
  {\dir/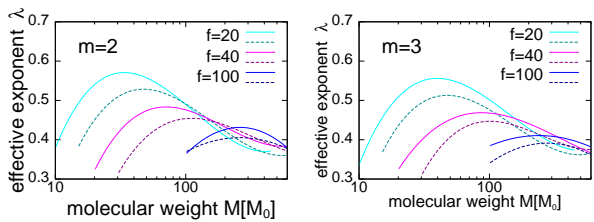}
\caption{
\new{''Effective exponents'' (differential fractal dimensions)}
 $\lambda = \ud \ln(R_g)/\ud \ln(M)$ versus molecular weight for 
$m=2$ (left) and $m=3$ (right). See text for explanation.
}
\label{fig:exponents}
\end{figure}

To quantify this further, we have fitted fourth order polynomials to the double
logarithmic data from the polymerization runs and used them to extract local
''effective exponents'', \new{{\em i.e.}, the differential fractal dimensions}
$\lambda(M) = \ud \ln(R_g)/\ud \ln(M)$. The results, shown in Fig.\
\ref{fig:exponents}, confirm the nonmonotonic behavior of the effective
scaling. The \new{differential fractal dimensions} initially increase and then
reach a peak with values that can be as large as $\lambda \sim 5.5$. The
position and the height of the peak depend on the molecular details: It moves
to higher values of $M$ and decreases if one increases $f$ or reduces the
reactivity of the monomers.  At large $M$, the \new{differential fractal
dimensions} decay and reach an asymptotic value which is slightly below
$\nu=0.4$. The asymptotic value is universal and does not depend noticeably on
$m$, $f$, and the monomer reactivity.

\begin{figure}
\includegraphics[angle=270, width=0.45\textwidth] {\dir/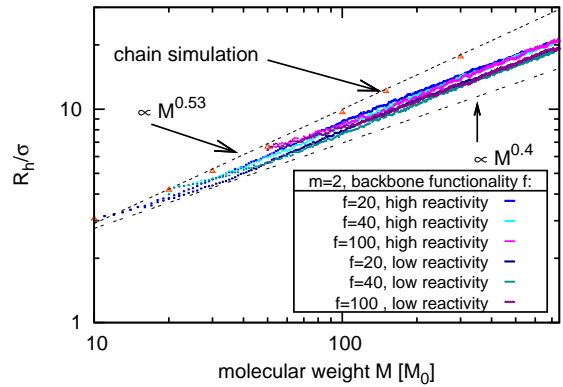}
\caption{
Hydrodynamic radius versus molecular weight (in monomer units) for 
monomer functionality $m=2$ various backbone core functionalities $f$, 
and the two different choices of reactivity, compared to simulation results 
for linear chains. Dashed lines indicate slopes corresponding 
to different scaling behavior. The results for $m=3$ are basically
identical.
}
\label{fig:hydrodynamic_radius}
\end{figure}

Next we discuss the hydrodynamic radius $R_h$, which is defined as the radius
of a Stokes sphere with the same diffusion coefficient as the polymer and
which we calculate {\em via} the approximate equation \cite{doi_edwards}
\new{
\begin{equation}
\frac{1}{R_h} = \frac{1}{N(N-1)} \sum_{i=1}^N \sum_{j \neq i} 
\frac{1}{|\vec{r}_i - \vec{r}_j|}.
\end{equation}
}
In the asymptotic limit, the scaling of $R_h$ and $R_g$ should be the
same. In the simulations, the results for $R_h$ are very similar to those 
for $R_g$, therefore we only show the data from polymerization runs for 
$m=2$ as an example (Fig.\ \ref{fig:hydrodynamic_radius}).
Much like in Fig.\ \ref{fig:gyration_1}, the data collapse onto two curves 
corresponding to high and low reactivity.
The two curves show a similar scaling behavior, which again 
differs distinctly from that of linear chains. We note that the
scaling of $R_h$ for linear chains does not quite reach the asymptotic 
limit $R_h \propto M^{\nu}$ with $\nu = 0.588$ in our system,
the apparent exponent ($\nu \approx 0.53$) is slightly smaller.
This is in agreement with results from other simulations and experiments
\cite{reith02}. In contrast, the scaling exponent of $R_h$ for the largest
hyperbranched chains, $\nu \approx 0.43(1)$, is slightly larger than that
measured for $R_g$.

\begin{figure}
\includegraphics[angle=0, width=0.45\textwidth] 
{\dir/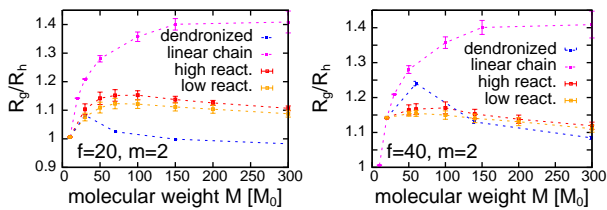}
\caption{
Ratio $R_g/R_h$ as a function of molecular weight in monomer units
for hyperbranched chains with monomer functionality $m=2$ and 
backbone functionality $f=20$ (left) and $f=40$ (right). Data for
linear chains and fully dendronized molecules (same backbone length)
are also shown for comparison.
}
\label{fig:Rg_Rh}
\end{figure}

The internal structure can be further investigated by inspecting the ratio
$R_g/R_h$. The value of $R_g/R_h$ depends on the molecular architecture. For
long ideal chains, it is known to be 1.5; experiments with linear polymers give
slightly smaller values \cite{bantle82}.  Dendrimers have been reported to have
ratios close to 1, star polymers with polydisperse arms have $R_g/R_h \approx
1.2$ \cite{burchard99,reith02b}.  Fig.\ \ref{fig:Rg_Rh} shows the data for
$R_g/R_h$ from equilibrium simulations for monomer functionality $m=2$ and
backbone length $f=20$ and $f=40$. Upon increasing the molecular weight, one
observes a distinct peak at $M \approx 50-100$ and a subsequent slight
decrease. A similar peak has been reported by Wang \etal \cite{wang11}
in simulations of hyperbranched polymers with a pointlike core. In our
si\-mulations, we observe that the peak is less pronounced for longer backbone 
(core)chains. It almost vanishes at $f=40$ and completely disappears at
$f=100$ (not shown). One can rationalize the peak as
follows: Starting from a linear chain, the polymer first develops into a star
shaped form with growing side chain number. With increasing size of side chains
and degree of branching, the overall structure becomes more spherical, \ie ,
$R_g/R_h$ decreases again. Dendronized molecules show a similar behavior
with an even more pronounced peak, whereas in linear polymers, $R_g/R_h$
grows monotonously as a function of $M$.

\begin{figure}
\includegraphics[angle=0, width=0.45\textwidth] 
{\dir/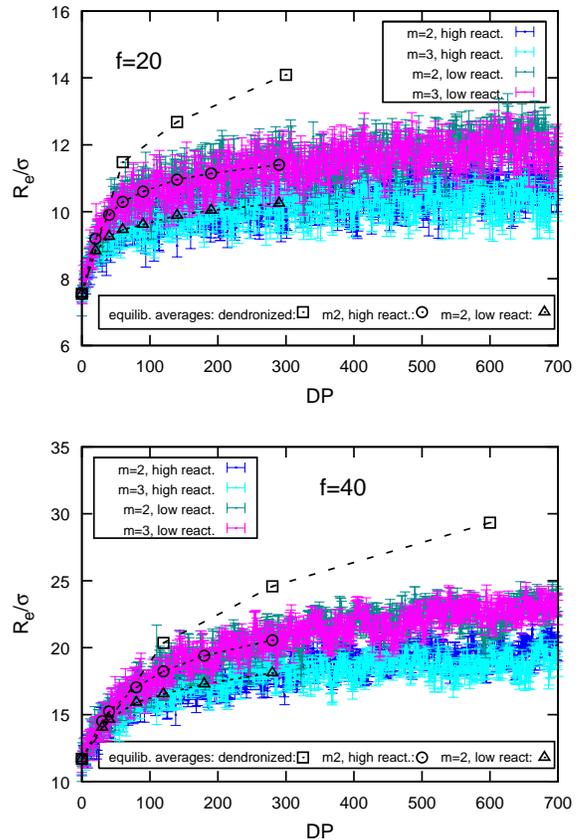}
\caption{
Average end-to-end distance of the backbone chain as a function of
the degree of polymerization (the number of hypergrafted monomers) for
backbone lengths $f=20$ (top) and $f=40$ (bottom). The black
symbols represent the averages from equilibrium simulations and 
the values from simulations of dendronized polymers.
}
\label{fig:stretching}
\end{figure}

Finally, we address the question whether and how the backbone chain stretches 
due to the growth of side chains. Naively, one might expect that the stiffness,
\ie, the persistence length of the backbone chain, can be tuned in a controlled
manner by hypergrafting branched side chains. However, extensive simulations of
bottle-brush polymers have indicated that the persistence length is no longer a
well-defined concept for such chains \cite{hsu10}. The same should hold for
linear-hyperbranched chains. Therefore, we will not discuss the stiffness of
the backbone here, but rather its elongation. The average end-to-end distance
$R_e$ is shown in Fig.\ \ref{fig:stretching} as a function of the number $\DP$
of grafted monomers for backbones of length $f=20$ and $f=40$ and monomer
functionality $m=2$. One can clearly distinguish two regimes: In the initial
regime up to $\DP \sim f$, where monomers attach directly to the backbone,
$R_e$ increases sharply with $\DP$. Then, the curves level off and
continue to increase much more slowly, such that the additional stretching of
$R_e$ relative to the end-to-end distance of the bare backbone is only of order
$10^{-4}$ per additional grafted monomer. We expect that this curve will
further flatten at much higher $\DP$ once $R_e$ becomes comparable to the
contour length, but this regime was not reached in the simulations. If monomers
have reduced reactivity, the range of the first regime is slightly extended
and $R_e$ increases. The slope of $R_e$ in the second high 
$\DP$-regime does not seem to depend on the monomer reactivity. 

For comparison, we have also studied the end-to-end distance of the 
corresponding dendronized molecules. The behavior in the initial regime 
is quantitatively similar to that of hyperbranched chains. In the second 
regime, the slope is much larger. This underlines once more the fundamental 
differences between dendronized molecules and hyperbranched molecules.

\section{Discussion and Conclusion}
\label{sec:conclusions}

We have studied the structural and conformational properties 
of linear-hyperbranched copolymers by computer simulations of a simple, 
solvent-free, coarse-grained spring-bead model. The model is designed to 
mimic the synthesis of linear-hyperbranched copolymers in good solvent 
under conditions of slow monomer addition. 
Our main results can be summarized as follows:

The topological properties of the hyperbranched chain (the chain
architecture) are intermediate between those of linear and dendronized chains.
%\haukeR{kann man das so sagen? bzw, in welchem sinne? irgendwie sind sie doch
%  auch eine eigene klasse von strukturen, zumindest verstehe ich die aussage
%  von \cite{juriju14} so, dass es naemlich bei den hyperbranched polymers
%  verschiedene universalitaetsklassen gibt.}
The evolution of 
the numbers of branching points with different coordination numbers
as a function total polymerization agreement is in very good agreement
to a simple rate theory proposed by H\"olter and Frey \cite{hoelter97b},
despite the fact that the theory neglects screening and excluded volume
effects. The theory predicts an upper limit for the degree of branching, 
hence our results suggest that the dendronized limit cannot be reached on
principle. The Wiener index $W$ is found to scale with the scaling exponent
$W \propto M^{\omega}$ with $\omega \approx 2.3$, which is between 
the exponents for dendrimers ($\omega = 2$ with logarithmic corrections)
and linear chains ($\omega = 3$). 

Likewise, the behavior of the polymer size (as specified by the radius of
gyration $R_g$ or the hydrodynamic radius $R_h$) as a function of molecular
weight is intermediate between that of linear and dendronized chains. The
scaling exponent is found to be around $\nu \approx 0.38$, closer to the
exponent of diffusion limited aggregation than to other exponents suggested in
the literature (1/2, 1/3, 1/4). However, the asymptotic regime might not have
been reached, and we cannot exclude the possibility that the true exponent is
$\nu = 1/3$ as predicted by Konkolewicz \etal \cite{konkolewicz07}.  At small
molecular weights, the exponent is nonmonotonic and varies between 0.4 and 0.5. 

We should note that these questions are not only of academic interest.
Molecular weight distributions of polymers are often measured by
chromatography, and they are calibrated by comparison with linear polymers of
known length. Thus one implicitly assumes that the hydrodynamic radii of the
target chains and the calibration chains behave similarly as a function of
molecular weight. If the exponent $\nu$ deviates significangly for both
architectures, the analysis becomes questionable and correction terms must be
applied. This requires a good knowledge of $\nu$ in the target chain.
Unfortunately, our results suggest that neither linear chains nor fully
dendronized chains are good reference systems. 

We have specifically addressed the question, whether molecules can be made
more compact by reducing the monomer reactivity. We find that reducing
the reactivity by a factor of three has relatively little influence on the 
degree of branching and no influence on the scaling properties of both
structural and conformational properties of the chains. It does, however,
influence the prefactor in the asymptotic power law for $R_g$ and $R_h$.
Hence the chains are more compact, but their behavior does not change
qualitatively.

We have also studied the distribution of monomers on the side chains and found
that it depends on the grafting point of the side chains on the core backbone.
For long backbone chains, the side chains that have polymerized at their ends
are about twice as long as the ones in the middle, leading to a pom-pom like
structure. This somewhat unexpected result of the present study will also have
implications for the synthesis and resulting properties of such structures.

Finally, we have examined the influence of the side
chains on the conformations of the backbone chain. As one would
expect, linear backbone chains stretch out if side chains are 
hypergrafted to them. However, the effect is much less pronounced than 
for dendronized chains.

Since our simulations are done with an implicit solvent model, they do not
include hydrodynamic interactions. In fluids at rest, hydrodynamics should
not be important on the time scales relevant for slow monomer
addition. This might change if one applies flow externally, and it might be
possible to manipulate the synthesis process by applying external flows, \eg,
in microfluidic setups. For example, linear-hyperbranched copolymers might
stretch in flow at high Weissenberg numbers, which should facilitate the
grafting of monomers to inner side chains. This will be the subject of future
investigations.

\acknowledgments{ 
We thank Christoph Sch\"ull for inspiring discussions. 
The simulations were carried out on the High Performance
Computer Cluster Mogon at Mainz university.
}

%\begin{appendix}
%\end{appendix}

%\bibliography{slabeof}% Produces the bibliography via BibTeX.

\begin{thebibliography}{99}

% Review

\bibitem{frey_book} H. Frey, C. Gao, D. Yan (Eds.)
 {\em Hyperbranched polymers: Syntheses, properties and applications},
 J. Wiley Publishers, 2011.

\bibitem{gao04} C. Gao, D. Yan,
% Hyperbranched polymers: From synthesis to applications
Progr. Polym. Science {\bf 29}, 183 (2004). 

\bibitem{voit09} B.I. Voit, A. Lederer,
% Hyperbranched and highly branched polymer architectures -
% synthetic strategies and major characterization aspects
Chem. Rev. {\bf 109}, 5924 (2009).

\bibitem{irfan10} M. Irfan, M. Seiler
% Encapsulation using hyperbranched polymers: From research and
% technologies to emergie applications
Ind. Eng. Chem. Res. {\bf 49}, 1169 (2010).

\bibitem{schuell13rev} C. Sch\"ull, H. Frey,
% Grafting of hyperbranched polymers: From unusual
% complex polymer topologies to multivalent surface
% functionalization
Polymer {\bf 54}, 5443 (2013).

% DENDRIMERS

\bibitem{bosman99} A.W. Bosman, H.M. Janssen, E.W. Meijer,
 %About Dendrimers: Structure, Physical Properties, and Applications
 Chem. Rev. {\bf 99}, 1665 (1999).

\bibitem{grayson01} S.M. Grayson, J.M.J. Frechet,
 %Convergent Dendrons and Dendrimers: from Synthesis to Applications,
 Chem. Rev. {\bf 101}, 3819 (2001).

\bibitem{gillies05} E.R. Gillies, J.M.J. Frechet,
 %Dendrimers and dendritic polymers in drug delivery
 Drug Discovery Today {\bf 10}, 35 (2005).

\bibitem{klos09} J. S. Kłos, J.-U. Sommer,
 %Coarse Grained Simulations of Neutral and Charged Dendrimers
 Polymer Science, Ser.C {\bf 55}, 125 (2013).

%%%%%%%%%%%%%%%%%%%%%%%%%%%%%%%%%

\bibitem{radke98} W. Radke, G. Litvinenko, A.H.E. M\"uller,
% Effect of core-formig molecules on molecular weight distribution
% and degree of branching in the synthesis of hyperbranched molecules
Macromolecules {\bf 31}, 239 (1998).

\bibitem{hanselmann98} R. Hanselmann, D. H\"olter, H. Frey,
% Hyperbranched polymers prepared {\em via} the core-dilution / 
% slow addition technique: computer simulation of molecular weight
% distribution and degree of branching
Macromolecules {\bf 31}, 3790 (1998).

\bibitem{bharathi00} P. Bharathi, J.S. Moore,
% Controlled synthesis of hyperbranched polymers by slow monomer addition
% to a core
Macromolecules {\bf 33}, 3212 (2000).

\bibitem{moeck01} A. M\"ock, A. Burgath, R. Hanselmann, H. Frey,
% Synthesis of hyperbranched aromatic homo- and copolyesters 
% via the slow monomer addition method
Macromolecules {\bf 34}, 7692 (2001).

\bibitem{schuell12} C. Sch\"ull, H. Frey,
% Controlled synthesis of linear polymers with highly branched side chains
% by hypergrafting: Poly(4-hydroxystyrene)-graf t-hyperbranched Polylycerol
ACS Macro Letters {\bf 1}, 461 (2012).

\bibitem{schuell13} C. Sch\"ull, H. Rabbel, F. Schmid, H. Frey,
% Polydispersity and molecular weight distribution of hyperbranched
% graft copolymers via ''Hypergrafting''of AB_m monomers from
% polydisperse macroiniator cores: Theory meets synthesis
Macromolecules {\bf 46}, 5823 (2013).

% RATE EQUATIONS

\bibitem{zimm49} B.H. Zimm, W.H. Stockmayer
%The dimensions of chain molecules containing branches and rings
J. Chem. Phys. {\bf 17}, 1301 (1949)

\bibitem{flory52} P.J. Flory,
J. Am. Chem. Soc. {\bf 74}, 2718 (1952).

% MEAN FIELD

\bibitem{lubensky79} T.C. Lubensky, J. Isaacson,
Phys. Rev. A {\bf 20}, 2130 (1979).

\bibitem{isaacson80} J. Isaacson, T.C. Lubensky,
% Flory exponents for generalized polymer problems
J. Physique Letters {\bf 41}, 469 (1980).

\bibitem{stauffer82} D. Stauffer, A. Coniglio, M. Adam,
Adv. Polym. Sci. {\bf 44}, 103 (1982).

\bibitem{cates85} M.E. Cates,
J. Physique {\bf 46}, 1059 (1985).

\bibitem{vilgis88} T.A. Vilgis, 
J. Physique {\bf 49}, 1481 (1988).

\bibitem{vilgis92} T.A. Vilgis, 
J. Physique II {\bf 2}, 2097 (1992).

\bibitem{buzza04} D.M.A. Buzza,
% Power law polydispersity and fractal structure of hyperbranched polymers,
Eur. Phys. J. E {\bf 13}, 79 (2004).

\bibitem{konkolewicz07} D. Konkolewicz, R. Gilbert, A. Gray-Weale,
% Randomly hyperbranched polymers
Phys. Rev. Lett. {\bf 98}, 238301 (2007)

\bibitem{konkolewicz08} D. Konkolewicz, O. Thorn-Seshold, A. Gray-Weale,
% Models for randomly hyperbranched polymers: Theory and simulation
J. Chem. Phys. {\bf 129}, 054901 (2008).

\bibitem{konkolewicz11} D. Konkolewicz, S. Perrier, D. Stapleton,
 A. Gray-Weale,
% Modeling highly branched structures: Description of the solution
% structures of dendrimers, polyglycerol, and glycogen,
J. Polym. Sci. Pol. Phys. {\bf 49}, 1525 (2011).


%SIMULATIONS

\bibitem{richards07} E.L. Richards, D.M.A. Buzza, G.R. Davies,
% Monte Carlo simulation of random branching in hyperbranched polymers,
Macromolecules {\bf 40} 2210 (2007).

\bibitem{wang10} L. Wang, X. He,
% Conformation of nonidal hyperbranched polymer in ABn (n=2,4)
% type polymerization
J. Polym. Sci. Pol. Phys. {\bf 48}, 610 (2010).

\bibitem{wang11} L. Wang, X. He, Y. Chen,
% Diffusion-limited hyperbranched polymers with substitution effect
J. Chem. Phys. {\bf 134}, 104901 (2011).

\bibitem{juriju14} A. Juriju, R. Dockhorn, O. Mironova, J.-U. Sommer,
% Two universality classes for random hyperbranched polymers
Soft Matter {\bf 10}, 4935 (2014).

\bibitem{lescanec90} 
\new{R. L. Lescanec, M. Muthukumar,
%Configurational characteristics and scaling behavior of starburst molecules:
%A computational study
Macromolecules {\bf 23}, 2280 (1990).}

%KONKOLEWICZ EXPERIMENTS

\bibitem{konkolewicz10} D. Konkolewicz, A. Gray-Weale, S. Perrier,
% Describing the structure of a randomly hyperbranched polymer
Macromolecular Theory Simul. {\bf 19}, 219 (2010).

\bibitem{konkolewicz07b} D. Konkolewicz, A. Gray-Weale, R.G. Gilbert,
% Molecular weight distributions from size separation data
% for hyperbranched polymers
J. Pol. Sci.: Part A {\bf 45}, 3112 (2007).

\bibitem{konkolewicz10b} D. Konkolewicz, A. Gray-Weale, S. Perrier,
% The structure of randomly branched polymers synthesized by
% living radical methods
Polym. Chem. {\bf 1}, 1067 (2010).

% PERFECTLY BRANCHED DENDRONIZED LINEAR CHAINS

\bibitem{schlueter00} D. Schl\"uter, J. Rabe,
% Dendronized polymers: Synthesis, characterization, 
% assembly at interfaces, and manipulation
Angew. Chem. Int. Ed.. {\bf 39}, 864 (2000).

\bibitem{zhang03} A. Zhang, L. Shu, Z. Bo, D. Schl\"uter,
% Dendronized polymers: Recent Progress in Synthesis
Macromol. Chem. Phys. {\bf 204}, 328 (2003).

\bibitem{frauenrath05} H. Frauenrath, 
% Dendronized polymers - building a new bridge from molecules to
% nanoscopic objects
Prog. Polym. Sci. {\bf 30}, 325 (2005).


% LINEAR HGC

\bibitem{lach98} C. Lach, R. Hanselmann, H. Frey, R. M\"ulhaupt,
Macromol. Rapid Commun. {\bf 19}, 461 (1998).

\bibitem{sun12} G. Sun, J. Hentschel, Z. Guan,
ACS Macro Lett. {\bf 1}, 585 (2012).

\bibitem{schuell12b} C. Sch\"ull, L. Nuhn, C. Mangold, E. Christ,
 R. Zentel, H. Frey,
 Macromolecules {\bf 45}, 5901 (2012).

%---------------------

\bibitem{wca} J.D. Weeks, D. Chandler, H.C. Andersen,
% Role of repulsive forces in determining the equilibrium structure
% of simple liquids
J. Chem. Phys. {\bf 54}, 5237 (1971).

\bibitem{espresso} H. Limbach, A. Arnold, B. Mann, C. Holm,
% ESPResSo: An extensible simulation package 
%  for research on soft matter systems
Comp. Phys. Comm. {\bf 174}, 704 (2006).

\bibitem{hoelter97} D. H\"olter, A. Burgath, H. Frey,
% Degree of branching in hyperbranched polymers
Acta Polymer {\bf 48}, 30 (1997).

\bibitem{hoelter97b} D. H\"olter, H. Frey,
% Degree of branching in hyperbranched polymers 2.
% Enhancement of the DB: Scope and limitations
Acta Polymer {\bf 48}, 298 (1997).


% WIENER INDEX

\bibitem{wiener47} H. Wiener,
% Structural determination of paraffin boiling points
J. Am. Chem. Soc. {\bf 69}, 17 (1947).

\bibitem{dobrynin01} A. Dobrynin, R. Entringer,
% Wiener index of trees; Theory and applications
Acta Applicandae Mathematicae {\bf 66}, 211 (2001).

\bibitem{bonchev92} D. Bonchev, O. Mekenyan, V. Kamenska,
% A topological approach to the modelling of polymer properties 
% (The tempo method)
J. Mathematical Chemistry {\bf 11}, 107 (1992).

\bibitem{widmann98} 
\new{A. H. Widmann, G. R. Davies,
%Simulation of the intrinsic viscosity of hyperbranched polymers
%with varying topology. Dendritic polymers built by sequential addition.
Comput. Theor. Polym. {\bf 8}, 191 (1998). }

\bibitem{sheridan02} P. Sheridan, D.B. Adolf, A.L. Lyulin, I. Neelov,
 G. D. Davies,
% Computer simulations of hyperbranched polymers: The influence of the
% Wiener index on the intrinsic viscosity and radius of gyration
J. Chem. Phys. {\bf 117}, 7802 (2002).

\bibitem{canfield85} E.R. Canfield, R.W. Robinson,
% Determination of the Wiener molecular branching index for the general tree
J. Comp. Chem. {\bf 6}, 7598 (1985).

\bibitem{note}  The scaling behavior of the Wiener index for fully
dendronized molecules can be estimated as follows: Molecules of generation $g$
contain \mbox{$N = f (m^g\!-\!1)/(m\!-\!1)$} monomers in total, and the number
of terminal monomers per side chain is $N_t = m^g$. In the limit $g \to
\infty$, one has $N_t \propto N$ and $g \propto \ln(N)$. Terminal monomers on
different side chains are separated by a path of length larger than $2g$, and
all paths are shorter than $2 g + f$. Thus we can estimate $2 g N_t^2 < W <
(2g+f) N^2$.  Since both the upper and lower bound of $W$ scale as 
$N^2 \ln(N)$, this must also hold for the Wiener index $W$ itself.  

\bibitem{gutman93} \new{I. Gutman, Y.-N. Yeh, S.-L. Lee, Y.-L. Luo,
%Some recent results in the theory of the Wiener number
Indian Journal of Chemistry {\bf 32a}, 651 (1993). }

\bibitem{polinska2014} 
 \new{P. Poli\`nska, C. Gillig, J. P. Wittmer, J. Baschnagel,
%Hyperbranched polymer stars with Gaussian chain statistics revisited
Eur. Phys. J. E {\bf 37}, 12 (2014)}

\bibitem{goetze03} I.O. G\"otze, C.N. Likos, 
% Conformations of flexible dendrimers: A simulation study
Macromolecules {\bf 36}, 8189 (2003).

\bibitem{giupponi04} G. Giupponi, D.M. Buzza, 
% Monte Carlo simulations of dendrimers in variable solvent quality
J. Chem. Phys. {\bf 120}, 10290 (2007).

\bibitem{doi_edwards} M. Doi, S. Edwards, {\em The theory of
polymer dynamics}, Oxford University Press, 1988.

\bibitem{reith02} D. Reith, 
% How does the chain extension of poly(acrylic acid) scale in aqueous
% solution? A combined study with light scattering and computer simulations
J. Chem. Phys. {\bf 116}, 9100 (2002).

\bibitem{bantle82} S. Bantle, M. Schmidt, W. Burchard,
% Simultaneous static and dynamic light scattering
Macromolecules {\bf 15}, 1605 (1982).

\bibitem{burchard99} W. Burchard,
% Solution properties of branched macromolecules
Adv. Polym. Sci. {\bf 143}, 113 (1999).

\bibitem{reith02b} D. Reith, M. Steinhauser, 
% Corrections to scaling in the hydrodynamic properties of dilute
% polymer solutions
J. Chem. Phys. 117, 914 (2002).


\bibitem{hsu10} H.P. Hsu, W. Paul, S. Rathgeber, K. Binder,
% Characteristic length scales and radial monomer density profiles
% of molecular bottle-brushes: Simulation and experiment
Macromolecules {\bf 43}, 1592 (2010).


\end{thebibliography}

\end{document}